\documentclass[aps,preprint]{revtex4}
\usepackage{amsmath}
\usepackage{graphicx}

\begin{document}

\title{Transport through ferromagnet/superconductor interfaces}

\author{Zhigang Jiang,$^1$ Jos\'{e} Aumentado,$^2$ Wolfgang Belzig,$3$ and Venkat Chandrasekhar$^1$}

\affiliation{$^1$Department of Physics and Astronomy, Northwestern University, 2145 Sheridan Road, Evanston, IL 60208, USA}
\affiliation{$^2$National Institute of Science and Technology, 325 Broadway, Boulder, CO  80305, USA}
\affiliation{$^3$Department of Physics and Astronomy, University of Basel, Klingelbergstr. 82, CH-4056 Basel, Switzerland}

\begin{abstract}
The differential resistance of submicron-size ferromagnet/superconductor interface structures shows asymmetries as a
function of the current through the ferromagnet/superconductor interface.  These asymmetries are a consequence of spin-polarized electron
transport from the ferromagnet to the superconductor, coupled with the Zeeman-splitting of the superconducting quasiparticle density of states. 
They are sensitive to the orientation of the magnetization of the ferromagnet, as the magnetic field required to spin-split the quasiparticle
density of states can be provided by the ferromagnetic element itself. 
\end{abstract}


\maketitle

\section*{Introduction}
There has been continuing interest in the past few years in heterostructures that combine ferromagnetic (F) and superconducting (S) elements.  This interest can be broadly divided into two categories: the superconducting proximity effect in ferromagnets, and spin-polarized electron transport between ferromagnets and superconductors.  In the first category are included earlier work on FS multilayers \cite{kawaguchi}, more recent work on critical currents in SFS junctions \cite{kontos}, and long-range superconducting proximity effects in ferromagnets adjacent to superconductors \cite{giroud,aumentado}.  Some of this work is discussed in other contributions in this volume.  In the second category is included spectroscopic studies of FS point contacts \cite{soulen}, extending the pioneering work of Tedrow and Meservey \cite{tedrow} in FS junctions mediated by an insulating tunnel barrier (I) to systems with higher transparency FS interfaces.  Point contact FS spectroscopy has been used to determine the degree of spin-polarization $P$ in the ferromagnet by examining in detail the differential conductance of the FS junction as a function of the voltage bias applied across it \cite{strijkers}, and analyzing the results in terms of a spin-polarized extension of the Blonder, Tinkham, Klapwijk (BTK) \cite{blonder} theory of transport across a normal-metal/superconductor junction.  Other theoretical studies have focused on the excess resistance of FS junctions associated with spin accumulation at the interface \cite{zhao,belzig}.

In the work of Tedrow and Meservey on FIS junctions, a magnetic field was applied to the thin film device in order to Zeeman-split the quasi-particle density of states in the superconductor.  The finite spin-polarization $P$ in the ferromagnet showed up directly as different peak heights in the FIS tunneling characteristics.  In the more recent work on point contact FS spectroscopy, a large magnetic field could not be applied, as this would rapidly suppress superconductivity in the bulk superconductors used in the experiments, so that the analysis of the current-voltage characteristics depends on a subtle interplay between $P$ and the FS interface transparency.  This restriction does not apply to mesoscopic FS devices made from thin films, where a magnetic field can be applied parallel to the plane of the superconducting thin film as in the original experiments of Tedrow and Meservey, splitting the quasi-particle density of states without substantially suppressing superconductivity in the film.  As in the FIS case, the finite spin-polarization in the ferromagnet is expected to show up as asymmetric peaks in the differential conductance as a function of voltage bias, as recently discussed by M\'{e}lin \cite{melin}.

In this paper, we report measurements of the low temperature differential resistance of mesoscopic FS junctions.  We observe asymmetries in the differential resistance even in the absence of an external magnetic field.  These asymmetries are associated with spin-polarized tunneling into the superconductor, with the splitting of the quasi-particle density of states in the superconductor arising from the magnetic field generated by the ferromagnetic elements.    

\section{Sample fabrication and measurement}
The samples for our experiments were fabricated by multi-level electron-beam lithography onto oxidized silicon substrates.  A number of samples corresponding to two different geometries were measured, but we present here results on only a few representative samples.  Figure 1(a) and (b) show scanning electron micrographs of our samples corresponding to the first and second types.  The first  device type (Fig. 1(a)) consists of an elliptical Ni particle in contact with a superconducting Al film.  The shape and size of the Ni particle, which was patterned and deposited first, ensures that it is single-domain, and that its magnetization lies along the major axis of the ellipse, as we have shown in previous experiments \cite{aumentado2}.  Four Au wires were then deposited to make contact with the Ni particle, and allowed us to make four-terminal measurements of the resistance of the Ni particle to characterize its electrical and magnetic properties.  The Al film was then deposited in the final lithography step.  Two of the Au probes were placed within 20-50 nm of the FS interface, enabling us to measure the resistance of the interface by itself, with very little contribution from the ferromagnet.  All interfaces were cleaned with an ac Ar$^+$ etch prior to deposition of the Au and Al films.  The thickness of the Ni films was $\sim$30 nm, and the Al and Au films $\sim$50-60 nm. Control samples of Ni, Al and Ni/Al interfaces were also fabricated in order to characterize the properties of the films and interfaces.  From low temperature measurements on these samples, the resistivity of the Ni film was estimated to be $\rho_{Ni} \sim6.6$ $\mu\Omega$cm, and that of the Al film $\rho_{Al} \sim 8.4$ $\mu\Omega$cm, corresponding to electronic diffusion coefficients $D_{Ni}\sim 76$ cm$^2$/s and $D_{Al}\sim 26$ cm$^2$/s respectively.  The second set of samples (Fig. 1(b)) were simple FS crosses, that enabled us to measure the four-terminal resistance of the FS interface directly.  In this second device type, both Ni and permalloy (NiFe) were used as the ferromagnet.

\begin{figure}[ht]
\center{\includegraphics[width=10cm]{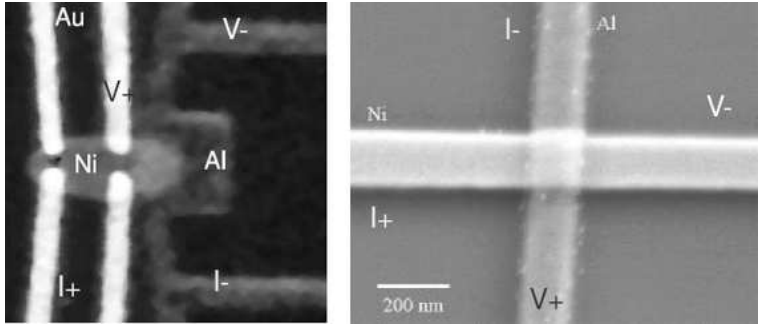}}
\caption{Scanning electron micrographs of representative device of the first type (a), and the second type (b).  (a) is scaled to 1$\mu$mx1$\mu$m.  The leads used to measure the four-terminal resistance are marked.  The current leads were used to send both the ac and dc current.}
\label{fig1}
\end{figure}
The measurements were performed down to 0.26 K in a $^3$He sorption insert with standard ac lock-in techniques.  The insert was placed in a dewar with a superconducting split-coil magnet, which allowed us to apply a magnetic field in the plane of the Ni particle, along its easy axis in the case of the first device type, and along the length of the ferromagnetic wire in the case of the second device type.  The application of the field in this direction has two advantages: first, the critical field of the superconductor is much greater than if the field were to be applied perpendicular to the plane of the sample, and second, the field direction is along the easy axis of magnetization of the ferromagnet.  We concentrate here on the differential resistance and conductance of the FS interfaces as a function of bias current; the temperature dependence of the resistance of these devices has been described in detail in an earlier publication \cite{aumentado}, and will not be discussed here.  In order to measure the differential resistance of a low resistance FS interface, we use a four-terminal configuration with the probe configurations as shown in Fig. 1, with a small ($\sim 10-50$ nA) ac current superposed on a dc bias current. 

\section{Experimental results and discussion}
Figure 2 shows the differential resistance of a permalloy/Al (Py/Al) cross similar to that shown in Fig. 1(b) at a temperature of $T=290$ mK, as a function of the dc current through the FS junction, at two different values of magnetic field.  As the current is increased from $\left|I_{dc}\right|$= 0, the resistance first increases, reaches a peak at a current of $\left|I_{dc}\right| \simeq$ 1 $\mu$A, and then shows two dips in the resistance before approaching the normal state resistance at higher values of $I_{dc}$.  The dips in $dV/dI$, corresponding to peaks in the differential conductance $dI/dV$, are asymmetric, in that the amplitude of the dips is different.  As the applied external field $H$ is increased, the features become \textit{sharper}, and more symmetric.  At larger values of $H$ (not shown in Fig. 2), the positions of the peaks and dips move down to lower values of $\left|I_{dc}\right|$.  Note that this behavior is strikingly different from what is observed in the FS point contact experiments.  Indeed, apart from the absence of the multiple peak structure, the resistance of the devices goes \textit{down} rather than up as one moves away from $I_{dc}$= 0.

\begin{figure}[ht]
\vspace{-0.2cm}
\center{\includegraphics[width=10cm]{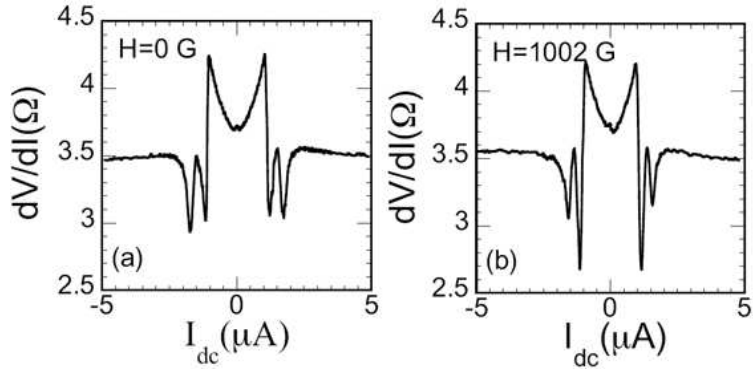}}
\caption{Differential resistance $dV/dI$ of a Py/Al cross, similar to the one in Fig. 1(b), as a function of dc current $I_{dc}$, at an external magnetic field of (a) $H=0$, and (b) $H=0.1002$ T.  $T=290$ mK.}
\label{fig2}
\end{figure}
In order to understand this behavior, we consider charge transport across the FS junction in the framework of a spin-polarized BTK model \cite{soulen,strijkers}.  In the conventional BTK model \cite{blonder}, the current across a NS junction with a voltage $V$ applied across it is given by
\begin{equation}
I_{NS}=2N(0) e v_F S \int_{-\infty}^\infty \left[f(E-eV)-f(E)\right]\left[1 + A(E) -B(E) \right] dE, 
\label{eqn1}
\end{equation}    
where $N(0)$ is the density of states at the Fermi energy, $v_F$ is the Fermi velocity of the electrons, $S$ is the cross-sectional area of the interface, $f$ is the Fermi function, and $A(E)$ and $B(E)$ are the BTK coefficients that represent the probability for Andreev reflection and normal reflection of electrons from the NS interface.  Their functional forms have been discussed in detail by BTK \cite{blonder}, and we shall not reproduce them here.  We only note that $A(E)$ and $B(E)$ depend on the transparency of the NS interface, characterized by the BTK parameter $Z$.  $Z$= 0 corresponds to a perfectly transparent interface, while $Z \rightarrow \infty$ corresponds to a tunnel barrier.  For $Z$= 0, $A(E)$= 1 for $E<\Delta$, the energy gap of the superconductor, and gradually reduces to zero as $Z$ increases.

The process of Andreev reflection involves a spin-up electron of energy $E$ coupling with a spin-down electron of energy $-E$ to form a Cooper pair in the superconductor.  If the normal metal is replaced by a ferromagnet with a finite polarization $P$, not all electrons of one spin species will be able to find a corresponding electron of the opposite spin species in order to form Cooper pair.  Hence, the probability of Andreev reflection will be reduced by a factor of (1-$P$), where we define the polarization $P$ by 
\begin{equation}
P= \frac{N_\uparrow(E_F) - N_\downarrow(E_F)}{N_\uparrow(E_F) + N_\downarrow(E_F)},
\label{eqn2}
\end{equation}  
$N_\uparrow(E_F)$  and $N_\downarrow(E_F)$ being the density of states for the up-spin electrons and down-spin electrons respectively \cite{p_note}.  The equation of BTK can then be modified to \cite{strijkers}
\begin{equation}
\begin{split}
I_{NS} = 2N(0) e v_F S & \int_{-\infty}^\infty \left[f(E-eV)-f(E)\right] \\
\big{(}(1-P)&\left[1 + A_u(E-g\mu H) -B_u(E-g\mu H) \right]  \\
           + P &\left[1 -B_p(E+g\mu H) \right]\big{)} dE,
\end{split}
\label{eqn3}
\end{equation}
where the original BTK coefficients $A(E)$ and $B(E)$ are modified to unpolarized ($A_u$, $B_u$) and polarized ($A_p$, $B_p$) versions, with $A_p$= 0.  This modification is required to maintain current conservation across the FS interface \cite{strijkers}, and involves a simple renormalization of the coefficients.  In addition to the formulation of Ref. \cite{strijkers}, we have also added the contribution to the energy of the quasiparticles from Zeeman splitting due to an external field $H$.  

Figure 3(a) shows the results of our calculations of the normalized differential resistance using Eqn.(\ref{eqn3}) with $H$=0, for a number of different combinations of $Z$ and $P$.  For $Z$=0 and $P$=0, we recover the usual BTK result, in that the resistance at zero bias drops to half the normal state value.  For $Z$=0 and $P$=0.5, the resistance at zero bias is exactly the same as the normal state resistance, while for $Z$=0.5 and $P$=0, it is slightly larger.   As both $Z$ and $P$ are increased, the resistance of the sample rises above the normal state resistance $R_N$, but $Z$ and $P$ affect the differential resistance in different ways at higher bias.  It is clear that the zero bias resistance is sensitive to both $Z$ and $P$, so one cannot determine both independently by examining only the zero bias resistance; the entire curve needs to be fit.  Figure 3(b) shows the results of similar calculations, but now with an applied magnetic field $H$= 0.2 $\Delta$.  The effect of the finite field is to cause a splitting in the structure of $dV/dI$ near $V\simeq\Delta/e$ due to the Zeeman effect.  The splitting can be seen even for $P$=0, but results in a substantial asymmetry of the curve when $P\ne0$.  

\begin{figure}[ht]
\center{\includegraphics[width=10cm]{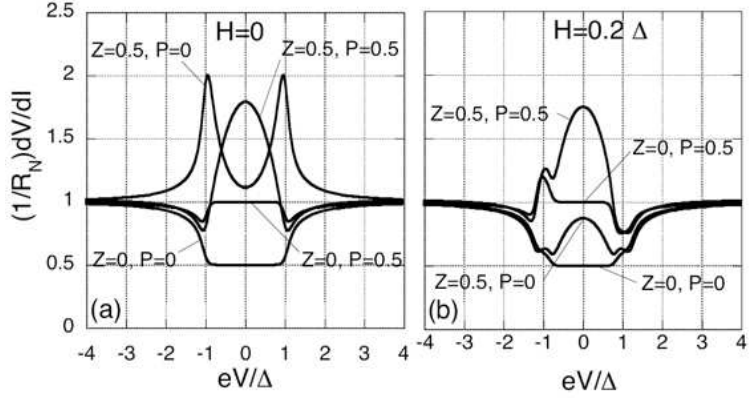}}
\caption{Numerical simulations of the differential resistance of a FS interface, from the theory described in the text.  (a)  $H$=0, (b) $H$=0.2 $\Delta$.  The values for $Z$ and $P$ are noted in the figure, and the temperature is $T$=$0.05\Delta$.}
\label{fig3}
\end{figure}

Comparing the results of these simulations to the data from the Py/Al cross shown in Fig. 2, we note that there are some similarities, but substantial differences in even the qualitative behavior.  First, the resistance in the simulations invariably decreases as the bias is shifted from 0, except for the case $Z$= 0.  In the experiment, the trend is opposite; the resistance \textit{increases} as the bias is shifted from 0.  This behavior is also reflected in the zero-bias resistance as a function of temperature in this sample, where a large increase in resistance is observed as one cools to below the transition temperature.  Second, at higher bias, two sharp dips appear in the data, which are not symmetric.  These are similar to the dips seen in Fig. 3(b), which are associated with Zeeman splitting of the quasi-particle density of states.  However, in the experiment, this structure is seen even at zero applied external field.  On the other hand, it should be noted that the field $H$ can result from a combination of the externally applied field and the self-field of the ferromagnet, which can be substantial near the ferromagnet.  At the superconductor, the direction of the externally applied field may be opposite that of the self-field of the ferromagnet, resulting in a decrease in $H$ as the external field is increased.  This apparently is the case for the sample of Fig. 2; as the external field is increased from 0 to 0.1 T, the two dips in $dV/dI$ become sharper and the spacing between them decreases, exactly as one would expect if the actual field at the superconductor decreased.  We have seen similar behavior in other devices as well.

\begin{figure}[ht]
\center{\includegraphics[width=8cm]{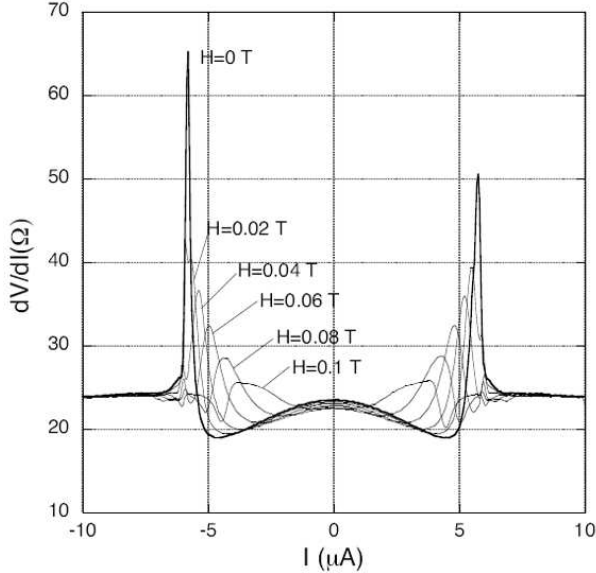}}
\caption{Differential resistance $dV/dI$ of the sample shown in Fig. 1(a), as a function of dc current $I_{dc}$, at different values of the applied magnetic field $H$.  The temperature $T=300$ mK.}
\label{fig4}
\end{figure}

The data shown in Fig. 2 was taken using the probe configuration shown in Fig. 1(b), and corresponds to a four-terminal measurement of the differential resistance of the FS interface alone.  Figure 4 shows the differential resistance $dV/dI$ of the sample of Fig. 1(a) as a function of dc current $I_{dc}$ at six different values of the external magnetic field $H$, applied along the major axis of the elliptical Ni particle, at $T$=300 mK.  Unlike the FS crosses, the four-terminal probe configuration used to measure this device (shown in Fig. 1(a)) includes a small portion of the Ni as well as the Al.  All the curves show an initial decrease in resistance as $I_{dc}$ is increased in either direction from zero, then an increase at $\left|I_{dc}\right|\simeq 5 \mu$A, and finally, all curves approach the normal state resistance $R_N$ at larger values of $\left|I_{dc}\right|$.  For low values of $H$, the peaks in $dV/dI$ are very sharp.  As $H$ is increased, the peaks decrease in amplitude, and move to lower values of $\left|I_{dc}\right|$.  In other devices we have measured, the peaks increase in amplitude and move to higher values of $\left|I_{dc}\right|$ as $H$ is increased from 0 initially, and then decrease in amplitude and move to lower values of $\left|I_{dc}\right|$ as $H$ is increased further.  This is simply due to the fact that the externally applied field can either add to or subtract from the field generated by the ferromagnetic element at the superconductor.  It should be noted that the peaks, especially at low magnetic fields, are not symmetric with respect to the applied current; the peak at negative current is higher than the peak at positive current.  This asymmetry is a manifestation of the injection of spin-polarized carriers into the superconductor.

The nature of the peaks is qualitatively different from those seen in the FS crosses, and in the simulations shown in Fig. 3, in that they are much sharper.  In fact, such peaks can be seen in the FS cross devices if one includes a portion of the superconductor in the four-terminal measurement.  Similar peaks have been observed in NS devices as well \cite{chien}, where they are associated with an excess resistance from a non-equilibrium charge imbalance in the superconductor.  Injection of quasiparticles into the superconductor results in a difference between the quasiparticle chemical potential $\mu_{qp}$ and the Cooper pair chemical potential $\mu_{cp}$.  $\mu_{cp}$ rises to its bulk value within a superconducting coherence length $\xi_S$ of the interface; $\mu_{qp}$ relaxes to $\mu_{cp}$ over a much longer length scale $\lambda_{Q^*}$, called the charge imbalance length.  In diffusive systems,  $\lambda_{Q^*}=\sqrt{D\tau_{Q^*}}$, where $D$ is the diffusion coefficient.  Near $T_c$, the charge imbalance time $\tau_{Q^*}$ is given by \cite{schmid}
\begin{equation}
\tau_{Q^*}=\frac{4 k_B T}{\pi \Delta(T,H)}\sqrt{\frac{\tau_{in}}{2 \Gamma}},
\label{eqn4}
\end{equation}
where $\tau_{in}$ is the inelastic scattering time, and $\Gamma$ is given by
\begin{equation}
\Gamma= \frac{1}{\tau_s} + \frac{1}{2 \tau_{in}}
\label{eqn5}
\end{equation}
and $\tau_s$ gives the contribution from orbital pair breaking.  The excess resistance arises from the difference between $\mu_{cp}$ and $\mu_{qp}$.  If the superconducting probe is placed a distance $x$ from the interface, an excess resistance $\Delta R$ will be measured, where $\Delta R=(\lambda_{Q^*}-x) \rho_S$, $\rho_S$ being the resistance per unit length of the superconductor.  $\lambda_{Q^*}$ diverges when $\Delta \rightarrow 0$, and this divergence gives rise to the peak in resistance seen in NS structures just below $T_c$.  

For a particle with an energy $E$, the effective gap seen is $\Delta -E$.  At low temperatures, we can take into account the excess resistance due to charge imbalance by introducing an effective \textit{voltage} dependent charge imbalance time $\tau_{Q^*}\simeq T/(\Delta-eV)$, and adding a resistance $R_{Q^*}$= $\rho_{Al}\lambda_{Q^*}$ to the resistance calculated in the spin-polarized BTK model described above.  

The solid line in Fig. 5 shows the result of our calculation including both spin-polarized BTK in a finite magnetic field as described above, and charge imbalance.  For comparison, we also show the result for only spin-polarized BTK.  These numerical calculations were performed by calculating the current through the FS interface for a particular voltage $V_0$, adding the charge imbalance voltage $V_{ci}$, and taking the derivative $d(V_0+V_{ci})/dI$ numerically.   The numerical calculations reproduce the large peaks seen in $dV/dI$ at currents corresponding to the gap, i.e.,  $I\simeq(\Delta/eR_N)$.  Unlike the BTK case, no splitting of the peaks near the gap is observed in the solid line in Fig. 5 (the dashed line, which shows the BTK case, does have a small splitting at negative $I$ that is difficult to observe because of the scale of the plot).  Closer examination of the position of the charge-imbalance peaks, however, shows that the peak at positive current is slightly greater than $\Delta/eR_N$, while the peak at negative current is slightly less than $-\Delta/eR_N$; the shift corresponds to the Zeeman splitting of the quasiparticle density of states.  Only one spin species gives rise to a peak for positive current; the other spin species gives rise to the corresponding peak for negative current.  Thus, the difference in the height of the peaks is an indication of the degree of spin polarization in the ferromagnet.  If we take the difference between the two peaks heights, divided by the sum of heights of the peaks taken from the normal state resistance, we obtain a value of 0.25, which is close to the value of $P=0.3$ used in the simulation.  A similar analysis performed on the experimental data of Fig. 4 gives a spin polarization of $P$=0.23, in good agreement with the expected value of the spin polarization for Ni.  

\begin{figure}[ht]
\center{\includegraphics[width=8cm]{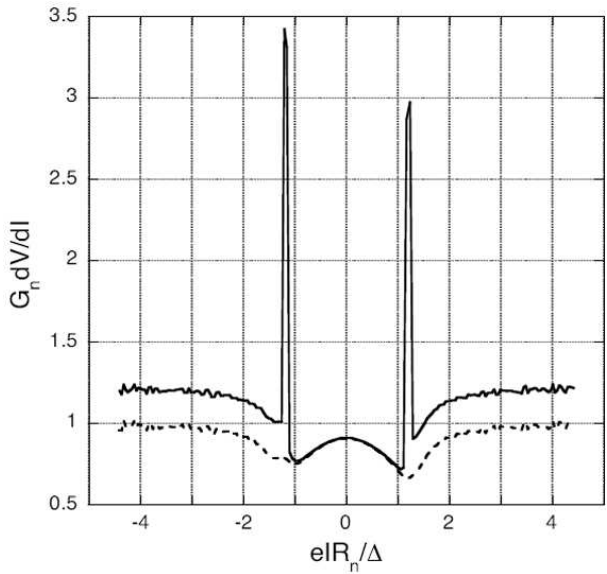}}
\caption{Numerical simulation of the differential resistance of a FS interface, including charge imbalance, as described in the text.  Solid line, spin-polarized BTK model with charge-imbalance; dashed line, spin-polarized BTK model without charge imbalance.  The other parameters used in the simulation are $P$=0.3, $Z$=0.3, and $H$=0.1 $\Delta$.}
\label{fig5}
\end{figure}

\section{Conclusions}
The differential resistance of mesoscopic ferromagnet/superconductor junctions shows a number of features associated with the injection of spin-polarized carriers into the superconductor.  In particular, large peaks are observed at currents corresponding to the superconducting gap voltage.  These peaks are not symmetric with respect to the current, in that their amplitudes are different.  The peaks are associated with the excess resistance arising from quasiparticle charge imbalance in the superconductor, and the difference in their heights is directly related to the degree of spin polarization in the ferromagnet. 

\begin{acknowledgments}
This work was supported by the US National Science Foundation under grant number DMR-0201530.
\end{acknowledgments}


\begin{thebibliography}{text}
\bibitem{kawaguchi}
K. Kawaguchi and M. Sohma, Phys. Rev. B \textbf{46}, 14722 (1992); J.S. Jiang \textit{et al.}, Phys. Rev. Lett. \textbf{74}, 314 (1995).

\bibitem{kontos}
T. Kontos \textit{et al.}, Phys. Rev. Lett. \textbf{86}, 304 (2001); V.V. Ryazanov \textit{et al.}, Phys. Rev. Lett. \textbf{86}, 2427 (2001).

\bibitem{giroud}   
M. Giroud \textit{et al.}, Phys. Rev. B \textbf{58}, R11872 (1998); M.D. Lawrence and N. Giordano, J.  Phys.: Condens. Matter \textbf{11}, 1089 (1999); V. Petrashov \textit{et al.}, Phys. Rev. Lett. \textbf{83}, 3281 (1999).

\bibitem{aumentado}
J. Aumentado and V. Chandrasekhar, Phys. Rev. B \textbf{64}, 54505 (2001).

\bibitem{soulen}
R.J. Soulen \textit{et al.}, Science \textbf{282} 85 (1998); S.K. Upadhyay \textit{et al.}, Phys. Rev. Lett. \textbf{81}, 3247 (1998). 

\bibitem{tedrow}
P.M. Tedrow and R. Meservey, Phys. Rev. Lett. \textbf{26}, 192 (1971); Phys. Rev. B \textbf{7}, 318 (1973).

\bibitem{strijkers}
G.J. Strijkers \textit{et al.}, Phys. Rev. B \textbf{63}, 104510 (2001); B. Nadgorny \textit{et al.}, Phys. Rev. B \textbf{63}, 184433 (2001); I.I. Mazin, A.A. Golubov, and B. Nadgorny, J. Appl. Phys. \textbf{89}, 7576 (2001); Y. Ji \textit{et al.}, Phys. Rev. Lett. \textbf{86}, 5585 (2001); C.H. Kant \textit{et al.}, Phys. Rev. B \textbf{
66}, 212403 (2002).

\bibitem{blonder}
G.E. Blonder, M. Tinkham, and T.M. Klapwijk, Phys. Rev. B \textbf{25}, 4515 (1982).

\bibitem{zhao}
H.L. Zhao and S. Hershfield, Phys. Rev. B \textbf{52}, 3632 (1995); F.J. Jedema \textit{et al.}, Phys. Rev. B \textbf{60}, 16549 (1999).

\bibitem{belzig}
W. Belzig \textit{et al.}, Phys. Rev. B \textbf{62}, 9726 (2000).

\bibitem{melin}
R. M\'{e}lin, Europhys. Lett. \textbf{51}, 202 (2000).

\bibitem{aumentado2}
J. Aumentado and V. Chandrasekhar, Appl. Phys. Lett. \textbf{74}, 1898 (1999).

\bibitem{p_note}
Another way to $P$ takes into account the Fermi velocities of the up spin and down spin electrons as well.  Since we use $P$ as a phenomenogical fitting parameter, the detailed definition of $P$ is not of concern here.

\bibitem{chien}
C.-J. Chien and V. Chandrasekhar, Phys. Rev. B \textbf{60}, 3655 (1999).

\bibitem{schmid}
A. Schmid and G. Sch\"on, J. Low Temp. Phys. \textbf{20}, 207 (1975).

\end{thebibliography}
\end{document}